\documentclass[prb,reprint,superscriptaddress]{revtex4-1}
\usepackage{H1}
\usepackage{amsmath}
\usepackage[pdftex,colorlinks=true]{hyperref}
\usepackage{amssymb}
\usepackage{amsthm}
\usepackage{amsfonts}
\usepackage{amsthm}
\usepackage{bbold}

\usepackage{bbm}
\usepackage{array}
\hypersetup{
citecolor = blue
}
\newcolumntype{L}[1]{>{\raggedright\let\newline\\\arraybackslash\hspace{0pt}}m{#1}}
\newcolumntype{C}[1]{>{\centering\let\newline\\\arraybackslash\hspace{0pt}}m{#1}}
\newcolumntype{R}[1]{>{\raggedleft\let\newline\\\arraybackslash\hspace{0pt}}m{#1}}

\newcolumntype{N}{@{}m{0pt}@{}}

\begin{document}
\title{Defining Time Crystals via Representation Theory}

\author{Vedika~Khemani}
\thanks{These authors contributed equally to this work.}
\affiliation{Department of Physics, Harvard University, Cambridge, Massachusetts 02138, USA}
\author{C.~W.~von~Keyserlingk}
\thanks{These authors contributed equally to this work.}
\affiliation{Department of Physics, Princeton University, Princeton, New Jersey 08544, USA}
\author{S.~L.~Sondhi}
\affiliation{Department of Physics, Princeton University, Princeton, New Jersey 08544, USA}

\begin{abstract}
Time crystals are proposed states of matter which spontaneously break time translation symmetry. There is no settled definition of such states. We offer a new definition which follows the traditional recipe for Wigner symmetries and order parameters. Supplementing our definition with a few plausible assumptions we find that a) systems with time independent Hamiltonians should not exhibit TTSB while b)  the recently studied $\pi$ spin glass/Floquet time crystal can be viewed as breaking a global internal symmetry and as breaking time translation symmetry, as befits its two names. 
\end{abstract}
\maketitle
\section{Introduction}
\label{sec:intro}
The identification of symmetries, and the categorization of phases through the spontaneous breaking of symmetries, is a central unifying theme in modern physics. This program has traditionally focused on the spontaneous symmetry breaking (SSB) of space-group and internal symmetries, and has provided a framework for understanding phenomena ranging from the formation of crystals to magnetism to superfluids. 

More recently, there has been much interest in the idea of \emph{time} translation symmetry breaking (TTSB) starting with Wilczek's proposal\cite{Wilczek12} for the existence of time crystals --- states of matter which spontaneously break continuous time translation symmetry down to a discrete subgroup, in direct analogy with spatial crystals.  Wilczek's intuition was sharpened by Watanabe and Oshikawa (WO) who formulated a  definition of TTSB in terms of space-time correlation functions, and used this definition to prove the absence of time crystals in thermal equilibrium\cite{Oshikawa15}. Interestingly, Ref~\onlinecite{Oshikawa15} left open the possibility of observing time crystals in an out of equilibrium setting. 

A parallel development\cite{Khemani15} showed that it is, in fact, possible to get non-trivial phase structure even in the intrinsically non-equilibrium setting of a periodically driven (``Floquet'') and disordered interacting system\cite{Lazarides14,Ponte15,Ponte15b,Abanin14,Rigol14} via a generalization of the notion of eigenstate order\cite{Huse13,PekkerHilbertGlass, Bauer13,Chandran14,Bahri15,Potter15} first formulated in the context of undriven many-body localized (MBL) systems\cite{Anderson58, Basko06, PalHuse, OganesyanHuse, Nandkishore14, AltmanVosk}.  Strikingly, one of the phases found in Ref.~\onlinecite{Khemani15} --- the so-called $\pi$ spin-glass phase --- was observed to display time-dependent correlations which oscillate at integer multiples $n$ of the driving frequency, thereby breaking the \emph{discrete} time-translation symmetry $\mathbb{Z}$ of the drive to a subgroup $n\mathbb{Z}$ and providing a realization of an out-of-equilibrium time crystal.  This phase (also variously known as the Floquet time crystal\cite{Q}/ discrete time crystal\cite{YaoDTC}) has been the subject of much recent study\cite{Khemani15, vonkeyserlingk16b, Q, CVS,YaoDTC, Q2}, and signatures of discrete TTSB in this setting have now been observed in very recent experiments\cite{MonroeTCExp,MishaTCExp}.

Three definitions of TTSB have been proffered. The first, following Watanabe and Oshikawa, defines a time crystal as a phase which shows time-dependence and long-range order in unequal space-time correlation functions of local operators\cite{Oshikawa15}
\begin{align} \label{eq:originalTTSB}
\lim_{|i-j|\rightarrow \infty}\lim_{L \rightarrow \infty} \langle O_i(t) O_j\rangle_c  = f(t), 
\end{align}
where $\langle\rangle_c$ denotes a connected correlation function, $L$ is the system size, $O_i$ is a local operator with support near position $i$, and $f(t)$ is a non-constant function of time. These correlators are measured in the Gibbs state or the ground state of an undriven Hamiltonian system. While one might think that the restrictions to large system sizes and large separations are not necessary to detect TTSB, in the absence of these restrictions essentially any finite system (or a chain of uncoupled finite systems) exhibits TTSB\cite{Oshikawa15}.  Ref~\onlinecite{Oshikawa15} also considered  
the natural generalization to correlation functions of superpositions of local operators $O = \frac{1}{L} \sum_i c_i O_i $, in which case TTSB requires $\lim_{L \rightarrow \infty} \langle O^\dagger(t) O\rangle_c = f(t)$.  
Generalized to the periodically driven setting\cite{CVS}, the correlators are  
measured in the eigenstates of the Floquet unitary $U(T)$ which is the time-translation operator over one period $T$, and the time-dependence is only probed ``stroboscopically'', $t=nT$ for $n\in \mathbb{Z}$. The second definition was proposed for Floquet systems; it states that such systems
exhibit TTSB if all eigenstates of $U(T)$ show long range connected correlations for local operators measured at equal times\cite{Q}. 
The third definition requires the presence of persistent, non-trivial time-dependence in expectation values of local operators measured in late time states starting from generic short range correlated initial states\cite{CVS, Q}: $$\lim_{t\rightarrow \infty} \lim_{L\rightarrow \infty} \langle\psi_0| O_i(t) |\psi_0\rangle =f(t),$$ where $f(t)$ is a non-constant function of continuous/stroboscopic time for Hamiltonian/Floquet systems. This last definition is appealing from an experimental viewpoint since neither the operator nor the initial state require fine-tuning; indeed recent experiments \cite{MonroeTCExp, MishaTCExp} have detected signatures of Floquet time-crystal states via an observation of multiple period oscillations in generic observables at late times.

In this work we consider a fourth definition of TTSB, which has the appeal of being a straightforward extension of standard definitions of SSB. We begin with a formal restatement of what SSB means for regular on-site global symmetries in terms of long range order in correlation functions of local order parameters.  We then discuss TTS much like any other symmetry in terms of its unitary implementation on Hilbert space and its action on the algebra of local observables.  This leads us to generalize the notion of an order parameter to TTS, where it becomes an observable which is fixed --- up to a non-trivial phase --- under the action of TTS. While the existence of local order parameters is guaranteed for on-site global symmetries, it is not at all guaranteed for TTS; the crucial difference is that TTS is typically not on-site, and typically spreads local operators. Indeed, in ergodic/ETH systems, no such order parameter exists because all local operators spread, precluding the existence of time crystals in such systems. Similarly, such order parameters do not exist in undriven MBL systems. Barring the presence of new dynamical phases, and some pathological unstable cases which we discuss, the only remaining possibility is Floquet-MBL phases where indeed examples of time-crystals have recently been found in theory\cite{Khemani15,vonkeyserlingk16b,Q,CVS,YaoDTC}, and possibly in experiments\cite{MonroeTCExp,MishaTCExp}. We comment on how the previous three definitions of TTSB can be deduced from the present one.

\noindent
\section{Standard lore on spontaneous symmetry breaking}\label{sec: standard SSB}

Consider a symmetry group $G = \{g\}$ represented by unitary operators $W(g)$ in the Hilbert space of our system\footnote{We restrict to unitary symmetries for this discussion.}.
For a Hamiltonian system $\mathcal{H}$ invariant under $G$, $[\mathcal{H},W(g)]=0 \; \forall \; g$,  the eigenstates $|n \rangle $ of $\mathcal{H}$ transform according to irreducible representations of $G$ (for finite systems of any size). As we later focus on situations where $G$ implements time translations, we limit our analysis to abelian groups $G$. For such groups,  all irreducible representations (irreps) are one-dimensional, and eigenstates simply pick up phases when acted upon by the symmetry group elements: $W(g)|n \rangle = e^{i\phi_g}|n\rangle$.

Traditionally, one considers two kinds of symmetry groups: (i) global internal onsite symmetries  and (ii) space group symmetries like translation and rotation. The unitary representations of the former are local \emph{i.e.} they are depth one quantum circuits which act as products of onsite unitary transformations. On the other hand, the unitary representations of spatial symmetries are non-local in that they cannot be represented as finite-depth circuits of local unitaries\cite{Po2016}. Relatedly, order parameters for detecting SSB (defined below) can always be chosen to be strictly local for internal symmetries, while they are forced to be extensive superpositions of local operators for spatial symmetries like translation. 

The case of time-translation symmetry is intermediate between these two cases. Time translations generated by local bounded Hamiltonians acting for finite times \emph{are} represented by local finite-depth circuits, but these are typically \emph{not} product circuits of on-site operators\footnote{Thus, perhaps counterintuitively, time translation symmetry is more similar to the case of a global symmetry rather than spatial translation symmetry.}. Correspondingly, as we will show, one can get order parameters for TTS which are either local operators (LOPs), or global superpositions of local operators (SLOPs).  Furthermore, there are systems where no local (or SLOP) order parameter for TTS can be defined at all, which in turn precludes the existence of TTSB in such systems. 

For concreteness, let us start by defining SSB for the case of global internal symmetries. We apologize in advance for our pedantry in the interests of a self-contained discussion. 
Consider a group $G$ represented by a product of onsite unitary operators $W(g)\equiv \otimes_i W_i(g) \;\forall\; g\in G$.
To define SSB from $G$ to a subgroup $H = \{h\}$, we construct a family of local operators $\{\Phi_{i, \alpha}\}$ labelled by position $i$, which transform according to irreps labelled $\alpha$ of $G$
\be
W^{\dagger}(g) \Phi_{i,\alpha} W(g) = e^{i \theta_{g,\alpha }}  \Phi_{i,\alpha}\punc{.}
\label{eq:op}
\ee
It is always possible to construct a local onsite basis of operators which transform in this way, due to the assumption that the symmetry acts as a product of single-site terms. For $\alpha$ non-trivial, the transformation of $ \Phi_{i,\alpha}$ under $G$ implies that one-point functions such as $\langle n |  \Phi_{i,\alpha} |n\rangle$ must vanish in all eigenstates $|n\rangle$. Given any $\alpha$ there is a unique irrep $\overline{\alpha}$ (the complex conjugate irrep) such that products $\Phi_{i,\alpha} \Phi_{j,\overline{\alpha}}$ transform trivially under $G$.  For the symmetry $G$ to be spontaneously broken to a subgroup $H$ in eigenstate $|n \rangle$, two conditions need to hold. (i) We must have
 \begin{align}
\lim_{|i-j|\rightarrow\infty}\lim_{L\rightarrow\infty}\left|\langle n|\Phi_{i,\alpha}\Phi_{j,\overline{\alpha}}|n\rangle_c 
\right| \neq 0 
\label{eq:SSBDef}
\end{align}
for all $\Phi_\alpha$ which transform trivially under $H$ but non-trivially under $G$, and (ii)  the same quantity \eqnref{eq:SSBDef} disappears for any $\Phi_\alpha$ which transforms non-trivially under $H$.

{\it Ising symmetry}: Let us consider a spin-$1/2$ system with a global Ising symmetry $G = \mathbb{Z}_2$. The degrees of freedom on each site $i$ can be represented by Pauli matrices $\sigma_i^{x/y/z/0}$ and the Ising symmetry is represented by $G = \{\mathbb{1}, P_x\}$ where $P_x = \prod_{i} \sigma_i^x$.  The transverse field Ising model $\mathcal{H} = J \sum_{\langle ij\rangle}  \sigma_i^z \sigma_{j}^z + h \sum_i\sigma_i^x$ is an example of such a system. The choice $\Phi_i = \sigma_i^z$ satisfies $P_x \Phi_i P_x = - \Phi_i$ so there exist local operators which transform under a non-trivial irrep of $G$. As before, all eigenstates can be chosen to be eigenstates of $P_x$. 
We say that Ising symmetry is spontaneously broken to $H = \{\mathbb{1}\}$ in eigenstate $|n\rangle$ if $\lim_{|i-j|\rightarrow \infty} \lim_{L \rightarrow \infty} \langle n| \sigma_i^z \sigma_j^z|n \rangle_{\text{c}} \neq 0$, where $\langle\rangle_{\text{c}}$ denotes a connected correlation function. 

In the following, we will largely frame our discussion of TTSB in terms of local order parameters. We will return to the case of SLOP(py) order parameters in Section~\ref{sec:SLOP}, where we also provide a comparison between the action of spatial and time translation symmetries. 
\section{Spontaneous time translation symmetry breaking (TTSB)}

\subsection{States, Operators and TTS}

Before defining TTSB for time translation, let us set up some notation for the two possible time translation groups $\mathbb{R},\mathbb{Z}$ corresponding to continuous and discrete time translation symmetry respectively, and consider the action of TTS on states and operators of Hamiltonian and Floquet systems. 

\subsubsection{Time independent Hamiltonian systems}  The group of time translations is the non-compact, abelian, Lie group $\mathbb{R}$ generated by the Hamiltonian itself. We note that TTS is therefore a \emph{dynamical} or Hamiltonian dependent symmetry rather than a \emph{kinematical} symmetry which would be the same for the entire set of Hamiltonians at issue\footnote{See e.g. Ref.~\onlinecite{PaisRMP}.}. The group elements are translations of the system by time $t$ with unitary representations $U(t)=e^{i \mathcal{H} t}$, and clearly $[\mathcal{H},U(t)]=0$ for all $t$. We set $\hbar=1$ throughout. Because the group is abelian, all irreducible representations  of TTS are one dimensional and eigenstates $|n\rangle$ of $\mathcal{H}$ transform in these irreps
\begin{align*}
U(t) |n \rangle = e^{-i E_n t} |n\rangle.
\end{align*}
We now turn to transformation of operators under conjugation with the group elements with a view towards defining an order parameter for TTSB. First, observe that it is certainly possible to find operators that transform according to the irreps of TTS. The operators $O_{nm}=|n\rangle \langle m|$ where $|n\rangle, |m\rangle$ are eigenstates of $\mathcal{H}$ with eigenvalues $E_n, E_m$ respectively, transform as
\begin{align}
\label{eq:TTSorderparameter}
U^\dagger(t) O_{nm} U(t) = e^{i (E_n - E_m)t} O_{nm} \equiv e^{i \Delta_{mn} t} O_{nm}
\end{align}
and form a basis for the linear space of all operators. While these operators certainly transform according to non-trivial irreps for $E_n \neq E_m$,   they are highly non-local and cannot, in general, serve as order parameters (which must be LOPs or SLOPs). Specifically, at issue is whether a subset of linear combinations of operators $O_{mn}$ are sufficiently local to serve as order parameters. Since we will require this order parameter to transform under a non-trivial irrep, we are only allowed to make linear combinations of basis-operators with the same $\Delta_{mn}$. In a generic many-body system with no additional symmetries, we do not expect degeneracies in the eigenvalue spacings so constructing such a local order parameter will, in general, not be possible. Below we return to the issue of when such local order parameters exist.

\subsubsection{ Periodic in time Floquet systems} For our purposes Floquet systems are characterized by local unitaries $U(T)$ which govern the evolution over a single period $T$. For such systems, the group of time translations is the infinite, abelian, discrete group $\mathbb{Z}$ generated by $U(T)$ itself. We note that this is again a dynamical rather than a kinematical symmetry. All irreps of this discrete TTS are one dimensional and eigenstates of $U(T)$  transform in these irreps
$$
U(T) |n\rangle = e^{-i \epsilon_n T} |n\rangle. 
$$
The $\epsilon_n$ are called ``quasienergies'' since they are only defined modulo $2\pi$. Again, the non-local, operators  $O_{nm}=|n\rangle \langle m|$ transform as
\begin{align}
U(T)^\dagger O_{nm} U(T) = e^{i (\epsilon_n - \epsilon_m)T} O_{nm} \equiv e^{i \Delta_{mn} T} O_{nm}
\label{eq:Omn}
\end{align}
and form a basis for the linear space of all operators. We defer again the question of the existence of local operators that transform via the irreps and note that we will not find cases where all do. 

\subsection{Defining TTSB via Local order parameters}
\label{sec:TTSBDef}

Let us adapt the previously discussed diagnostic of SSB to the case of TTSB. We require the existence of a family of local order parameters $\Phi_{i,\alpha}$ that transform under non-trivial irreps of the TTS ($\mathbb{R}$ or $\mathbb{Z}$) such that $U^\dagger(t) \Phi_{i,\alpha} U(t) = e^{i\Delta_{\alpha} t} \Phi_{i,\alpha}$. Then, it follows that $\langle n| \Phi_{i,\alpha} |n\rangle=0$ in every eigenstate (where $|n\rangle$ is an eigenstate of either $\mathcal{H}$ or $U(T)$ depending on the context, and correspondingly $t\in \mathbb{R},\mathbb{Z}T$ respectively). A system breaks TTS to a discrete subgroup $H$ (of $\mathbb{R}$ or $\mathbb{Z}$), if the following two conditions hold. (i)  We have
\begin{align}
\lim_{\left| i-j\right|\rightarrow\infty}  \lim_{L \rightarrow \infty} \left|\langle n| \Phi_{i,\alpha} \Phi_{j,\overline{\alpha}} |n\rangle_{\text{c}} \right| =c_0 \ne 0.
\label{eq:TTSBDef}
\end{align}
for $\Phi_\alpha$ which transform trivially under $H$ and non-trivially under $G$. (ii) The same correlator disappears for $\Phi_\alpha$ which transform non-trivially under $H$.

\section{When does TTSB occur?}
Having defined TTSB, we consider the scenarios under which TTSB could potentially occur in the infinite volume limit. We will see that thermalizing systems and undriven MBL phases are precluded from showing TTSB because, on general grounds, they do not posses a local TTS order parameter, as pre-empted in the discussion below \eqnref{eq:TTSorderparameter}. Saving the possibility of exotic phases which are neither localized nor thermalizing, the only remaining known phases of matter are Floquet MBL phases; hence if TTSB phases arise, they must do so in such systems.

\subsection{ Absence of TTSB for thermalizing systems} For generic, thermalizing interacting systems, it is generally believed that any local operator spreads ballistically under time evolution at a characteristic Lieb-Robinson velocity $v_{\text{LR}}$\cite{Lieb72}. This spreading can also generically be sub-ballistic, such as in thermalizing disordered systems with Griffiths effects\cite{AgarwalGriffiths}. 
Hence, thermalizing systems lack any candidates for TTS order parameters which are local (or SLOPs) and thus they cannot exhibit TTSB. As a consequence of this operator spreading, we expect all time dependent correlations of the form $\langle n | O_i(t) O_j |n \rangle$ for generic local operators $O_i$ to asymptote to constants at long times. This is true of both Hamiltonian systems and Floquet systems. In the latter, the ergodic phase is the infinite temperature phase since it is believed that a generic, interacting driven system absorbs energy indefinitely from the drive in the absence of localization, and thus all late-time correlations are trivial in the Floquet-ergodic setting\cite{Lazarides14PRE,Rigol14,Abanin14,Ponte15}. 

\subsection{Absence of TTSB for Hamiltonian MBL systems} Hamiltonian systems which are localized (either MBL or Anderson localized) have an extensive set of mutually commuting local operators --- called ``l-bits''\cite{Serbyn13a,Serbyn13cons,Huse14,Ros15,Chandran15b}--- which commute with each other and with $H$. 
These local l-bit operators certainly transform according to an irrep of TTS, but it is the trivial irrep because of their commuting with $\mathcal{H}$. As a result, they cannot serve as order parameters for diagnosing TTS. On the other hand, local operators which do not commute with $\mathcal{H}$ grow logarithmically in time\cite{Serbyn13a,Serbyn13cons,Huse14,Swingle16,Huang16,Fan16,Chen16} and again cannot serve as order parameters for TTS. Hence  TTSB is not possible for such systems either. 

\subsection{Presence of TTSB for Floquet MBL systems} Having excluded the possibility of TTSB in thermalizing and undriven MBL systems, we now turn to the category of systems that do exhibit TTSB. We will show that one can define local order parameters for TTSB in Floquet MBL systems and, in such systems,  time-translation symmetry breaking is accompanied by the breaking of a global (kinematical or dynamical) internal symmetry. Our primary example is the one dimensional perturbed $\pi$ spin-glass phase described by the family of model unitaries\cite{Khemani15,vonkeyserlingk16b,CVS,Q}
\begin{align}
U(T) = P_x \exp\left[{i \sum_{j} J_{j} \sigma_j^z \sigma^z_{j+1} +  h^x_j \sigma^x_j + h^y_j \sigma^y_j + h^z_j \sigma^z_j} \right]
\label{eq:UfMBL}
\end{align}
where $P_x=\prod_j \sigma^x_j$ is the Ising parity operator. For sufficiently disordered fields $h^{\alpha}_j$ and couplings $J_j$, the system is many-body localized.  We direct the reader to Ref~\onlinecite{CVS} for more technical details on this model. 
We consider several cases below. 

\subsubsection{$h^x_i= h^y_i = h^z_i= 0$}
Let us start with the special choice $h^x_i= h^y_i = h^z_i= 0$. For this case,  $U_0(T) = P_x e^{i \sum_j J_j \sigma^z_j \sigma^z_{j+1}}$  has a global $\mathbb{Z}_2$ Ising symmetry, $[U_0(T),P_x]=0$. Note that 
\begin{align}
U_0^\dagger(T) \sigma^z_i U_0(T) = -\sigma^z_i, \nonumber\\
U_0^\dagger(2T) \sigma^z_i U_0(2T) = \sigma^z_i
\label{eq:PSGop}
\end{align}
whence $\sigma^z_i$ is a local operator which transforms under a non-trivial irrep of TTS and can serve as an order parameter for diagnosing discrete TTSB from $G = \mathbb{Z}$ to the subgroup $H = 2\mathbb{Z}$ \eqref{eq:op}. It is, of course, {\it also} an order parameter for the $\mathbb{Z}_2$ Ising symmetry generated by $P_x$. 

The eigenstates of $U_0(T)$  look like Ising symmetric/antisymmetric global superposition states (cat states) of frozen $\sigma^z$ spins\cite{Khemani15, vonkeyserlingk16b, CVS} , $|\pm\rangle = \frac{1}{\sqrt{2}} \left(| \uparrow \downarrow\downarrow \cdots \uparrow\rangle \pm | \downarrow \uparrow\uparrow \cdots \downarrow\rangle\right)$.
Two point functions of the order parameter show long-range order, $\langle \pm| \sigma_i^z \sigma_j^z |\pm \rangle_c = \pm 1$ in all eigenstates. Thus, 
all eigenstates spontaneously break \emph{both} the discrete TTS $\mathbb{Z}$ down to $2\mathbb{Z} $ \emph{ and} the global Ising $\mathbb{Z}_2$ symmetry; both of these breakings are diagnosed by the same order parameter---whence the original designation of this spatiotemporally ordered\cite{CVS} phase as a``$\pi$ spin-glass''\cite{Khemani15}. 

Furthermore, it is easy to check that the $|+\rangle, |-\rangle$ eigenstate pairs for a fixed configuration of spins differ in quasienergy by $\pi/T$.  This $\pi/T$ quasienergy splitting between the constituents of each feline doublet\cite{Khemani15, CVS} is essential for defining a local order parameter for TTS. Referring back to our general arguments around \eqnref{eq:TTSorderparameter}, we see that this system has $2^{L-1}$ independent (non-local) basis operators $O_{\pm} =|+ \rangle\langle -  | $ labeled by the domain wall configurations which all transform in the same non-trivial way as $O_{\pm}(T) = - O_{\pm}$ with $\Delta_{+-} = \pi/T $ (\emph{c.f.} Eq.~\eqref{eq:Omn}).
This degeneracy allows us to construct local operators $\sigma^z_i$ which themselves transform in the same way by taking superpositions of these basis operators.  
\subsubsection{$h^x_i\neq 0,\; h^y_i \neq 0,\; h^z_i\neq 0$}

Consider the general case where all three fields $h_i^{x/y/z}$ are small but non-zero. Note that this model does \emph{not} have a global Ising symmetry $P_x$. Moreoever, the $\sigma_i^z$ no longer transform according to irreps of $U(T)$ so they cannot serve as order parameters for diagnosing TTSB. 
However, if the couplings are sufficiently random so that the system remains MBL, there is a finite depth unitary circuit $V$ which relates the eigenstates of the perturbed unitary to those of the unperturbed model $U_0(T)$, and which can be used to define dressed, drive-dependent, local l-bit operators\cite{CVS} $\tau_i^\alpha =  V \sigma_i^\alpha V^{\dagger}$. For sufficiently large system sizes the unitary can be rewritten in a canonical form\cite{CVS} 
\begin{align}\label{eq:lbittimecrystal}
U(T) = \tilde{P}_x e^{i D_{\rm even}({\tau^z_i})} \punc{,}
\end{align}
where $\tilde{P}_x=\prod_i \tau_i^x$ and $D_{\rm even}$ is an even function of the $\tau_i^z$ operators. Notice that this unitary has an emergent, dynamical global $\mathbb{Z}_2$ symmetry generated by $\tilde{P}_x$. 
Evidently the $\tau_i^z$ are  local order parameters for both TTSB \emph{and} the global $\mathbb{Z}_2$ generated by $\tilde{P}_x$, which is now also a dynamical symmetry. The eigenstates of the unitary are still $\pi/T$-paired cat states of the dressed $\tau^z$ spins which are now labeled by their eigenvalues under the dressed parity operator. Thus, in this range of parameters, the system spontaneously breaks both the dynamical TTS and the dynamical Ising symmetry $\tilde{P}_x$ and both phenomena are detected by the \emph{same} order parameters. Upon exiting the phase in a general direction, order parameters for both symmetries can cease to exist\footnote{We note that for perturbed drives which preserve Ising symmetry ($h^y = h^z = 0, h^x\neq 0$), the $\mathbb{Z}_2$ symmetry breaking is still diagnosed by the $\sigma_i^z$ operators. On the other hand, order parameters for TTSB are defined in terms of dressed l-bit operators $\tau_i^z$. }. This discussion emphasizes that the spatial long-range order and time translation symmetry breaking in this model go hand in and hand and should be viewed on an equal footing\cite{CVS}.

This analysis can be generalized {\it mutatis mutandis} to drives with $\mathbb{Z}_n$ kinematical/dynamical symmetries described in Refs.~\onlinecite{vonkeyserlingk16b,CVS,Sreejith16}.

\section{TTSB vs. spatial translation symmetry breaking}
\label{sec:SLOP}
We now briefly discuss the existence of order parameters for TTSB which are global superpositions of local operators, and offer a comparison between TTSB and the spontaneous breaking of spatial translation symmetry. 
Let us consider discrete spatial translation symmetry $G = \mathbb{Z}$ in a 1D system with  $L$ sites. This symmetry cannot be represented by a finite-depth circuit of local unitary operators. For an operator $O(r)$ centered at site $r$, translation by $x$ sites acts as $\hat{T}^{\dagger}(x) O(r) \hat{T}(x)=O(r+x)$. This implies that no nontrivial local operator is left fixed by spatial translations. Thus, the closest analogue to a local order parameter (as defined in \eqnref{eq:op})
involves a global superposition of local operators
\begin{align*}
O_{k} & =\frac{1}{L} \sum_r \; e^{ikr}O(r)\\
\hat{T}^{\dagger}(x) O_k \hat{T} (x)  & =e^{- ik}O_{k}
\end{align*}
where the wavenumber is restricted so that $e^{i k L}=1$. Lattice translation symmetry can be broken down to a discrete subgroup $n\mathbb{Z}$ (as in an antiferromagnet). Through a direct generalization of  \eqref{eq:SSBDef}, this order can be diagnosed by non-trivial connected correlations in $\langle O_{-G} O_{G} \rangle_c$ where $G \in \frac{2\pi}{n}\{1,2,\ldots,n-1\}$. 

By contrast, time translation acts rather differently. Generically, there is no basis of operator eigenvectors for time translation which look like SLOPs. In generic ergodic phases, for example, operators which are eigenvectors under time translation involve highly non-local superpositions of large strings of operators. Even in MBL systems, {\it most} local operators grow logarithmically\cite{Serbyn13a,Serbyn13cons,Huse14,Swingle16,Huang16,Fan16,Chen16}  in time and do not transform as eigenoperators under time translation. Most eigenoperators of time-translation in MBL systems (all except the l-bits) again involve non-local superpositions of large strings of operators. 

There are, however, examples of fine-tuned local Clifford circuits called ``gliders''\cite{Gutschow10} for which one can construct order parameters which are global SLOPs, in direct analogy with spatial crystals. This is striking since these circuits, unlike spatial translation, are local and can be viewed as the time evolution operator generated by a local Hamiltonian. We defer the existence of TTSB in such glider circuits with SLOP order parameters, and the stability of these circuits to generic perturbations, as interesting questions for future work.

\section{Relation with prior definitions}
We now briefly discuss how our definition of TTSB makes contact with the prior three definitions mentioned in the Introduction. We will stick to the case of local order parameters for simplicity, although the generalization to SLOPs is straightforward. Definition 1 requires long-range order in unequal space-time correlation functions of local operators\cite{Oshikawa15} \eqref{eq:originalTTSB}. Indeed, if a system exhibits TTSB per our definition in Section~\ref{sec:TTSBDef}, then there exist local order parameter operators $\Phi_{i,\alpha} $ such that $\Phi_{i,\alpha}(t) = e^{i\Delta_{\alpha}t} \Phi_{i,\alpha} $. Using \eqnref{eq:TTSBDef}, we find the existence of oscillating spatiotemporal order as desired:
\begin{align}\label{eq:OPautocorr}
\lim_{|i-j|\rightarrow \infty } \lim_{L \rightarrow \infty} \left|\left\langle  n |\Phi_{i,\alpha}(t)\Phi_{j,\overline{\alpha}} | n \right\rangle_{\text{c}}\right|=  e^{i\Delta_{\alpha}t}c_{0}
\end{align}
for either $t\in \mathbb{R}$ or $t\in \mathbb{Z}T $.
Moreover, replacing the order parameters in \eqnref{eq:OPautocorr} with generic local operators $O_{i}$ will also lead to persistent oscillations at long times, although more than one frequency can be present -- this occurs because  the $O_i$ will generically have some overlap with $\Phi_{i,\alpha}$. 
We note that our formalism for TTSB looks for long range order in correlations of order parameters, so the limit $|i-j|\rightarrow \infty$ is natural. In the absence of this limit, unequal spacetime correlation functions of generic operators exhibit ``glassy'' temporal dynamics and oscillate with multiple incommensurate frequencies\cite{CVS}. Moreover, as mentioned earlier, in the absence of this limit even ``trivial'' systems like uncoupled chains of Rabi oscillators appear to break TTSB.

Of course, our definition requires the existence of long-range order in connected correlators of local order parameters  in Floquet eigenstates \eqnref{eq:TTSBDef}. Thus, such eigenstates fail to cluster and must be long-range correlated in agreement with the second definition of Floquet time crystals\cite{Q}. 

Finally, our definition is almost equivalent to the third definition in terms of quenches and late time states which is the most
experimentally useful one\cite{CVS}. The non-trivial transformation of the order parameter under TTS means that the order parameter is also an exact spectrum generating operator. Thus, its existence implies the existence of spectral multiplets separated in quasienergy by $\Delta_\alpha$ as discussed for the $\pi$-spin glass. This in turn implies \cite{CVS} that at late times starting from generic short range correlated initial state, generic local observables exhibit oscillations with a definite period $2\pi/\Delta_\alpha$---provided the system is MBL. The contributions from other frequencies dephase away as power laws of time as can be deduced from the l-bit formalism of localized systems\cite{CVS}. However, if the system lacks interactions and is Anderson localized, the late time state can also continue to exhibit multiple---indeed infinitely many---periods which are a consequence of the existence of an infinite number of exact local raising operators in the non-interacting problem.

\section{Concluding Remarks}  

In this work, we have presented a definition of TTSB by treating time translation much like any other Wigner symmetry in terms of its unitary implementation on Hilbert space, its action on the algebra of local observables and the consequent identification of order parameters, and the correlations of the order parameter in eigenstates. Local order parameters for TTS do not generally exist. When they do, the eigenstates show long-range connected correlations with respect to the order parameter. We further argue, with some minor assumptions, that the only setting for the existence of a TTS order parameter and hence TTSB is in Floquet MBL systems. 
This recovers, albeit heuristically, the result of Oshikawa and Watanabe for the absence of TTSB in equilibrium systems using their definition in terms of space-time 
correlators. It also provides an extension to the case of localized systems which are out of equilibrium and which cannot be obtained via their
methods. 
In Floquet MBL systems, the breaking of the dynamical time translation symmetry is always accompanied by breaking of a (kinematical or dynamical) global internal symmetry, befitting the varying descriptions of this spatiotemporally ordered phase\cite{CVS} as a $\pi$ spin-glass\cite{Khemani15,vonkeyserlingk16b}/Floquet time-crystal\cite{Q}.

{\it Acknowledgements:}
We would like to thank David Huse for many useful discussions on this topic. This work was supported by the NSF-DMR via Grant No. 1311781 (CvK and SLS), the Princeton Center for Theoretical Science (CvK), the Alexander von Humboldt foundation via a Humboldt award (SLS), and by the Harvard Society of Fellows (VK). 
\begin{appendix}
\section{Non-local order parameters}
While this paper was entirely about the experimentally relevant case of a local order parameter for TTSB, we note that it may be interesting
to explore theoretically the existence of drives which do not admit local order parameters but for which non-local operators transform as non-trivial irreps of TTS. One example noted 
{\it en passant} in Ref.~\onlinecite{CVS} is that of the following drive
\begin{align}
U(T) &= e^{i H_\text{TC}} \prod_{e \in C^*_1} \sigma^z_e \\
H_{\text{TC}} &= -\sum_p \epsilon_p \prod_{e\in\partial p} \sigma^x_e -  \sum_v \epsilon_v \prod_{e\in  s(v) } \sigma^z_e 
\end{align}
Here $H_{\text{TC}}$ is the standard toric code Hamiltonian\cite{Kitaev03}, involving a sum over plaquette $p$ and vertex $v$ terms. $C^*_{1,2}$ are collections of edges corresponding to closed paths on the dual lattice;  the two paths are non-contractible loops winding around the two directions of the torus. Consider the Wilson loop $W_{C_2}\equiv \prod_{e\in C_2} \sigma^x_e$ corresponding to a non-contractible loop on the lattice around the $y$ direction on the torus. This operator transforms non-trivially under TTS, i.e., $W_{C_2}(nT)=(-1)^n W_{C_2}$. Now the order parameter is non-local, but formally we can examine the expectation value of the product of two such Wilson loops at large separation. This does not vanish and signals both the presence of $\mathbb{Z}_2$ topological order and TTSB. 
It would be interesting to see if there are other more naturally symmetric examples of TTSB, once the constraint of locality is lifted.

\section{Yet another characterization of TTSB}
In the ground state (GS) of the transverse field Ising model, we  usually diagnose SSB through the presence of long range order $\left\langle \sigma^z_{r}\sigma^z_{s}\right\rangle _{\text{GS,c}}=O(1)$. But there is a dual diagnosis of this phenomenon using the disorder operator. In this language, long range order is signaled by the fact that the action of Ising symmetry on a subregion $R$ obeys a volume law
$$
\left\langle \prod_{r \in R}\sigma^x_{r}\right\rangle _{\text{GS}}\sim Ce^{- \alpha \text{vol}({R})}\punc{.}
$$
In contrast, in the paramagnetic phase, this correlator exhibits a boundary law $\sim e^{- \alpha |\partial{R}|}$. Motivated by this, and using the form \eqnref{eq:lbittimecrystal}, suggests an alternative dual characterization of time translation symmetry breaking in Floquet systems with local Hamiltonians $H(t)$. Define the local action of time translation in region $R$ as
$$
U_{\text{f},R}(T) = \mathcal{T}e^{-i \int^{T}_0 dt H_R(t)} \punc{,}
$$
where $H_R(t)$ are those terms in $H(t)$ which have support solely in $R$. With the notation setup, we conjecture the following for Floquet systems with TTSB:  If a Floquet system exhibits time translation symmetry breaking $\mathbb{Z}\rightarrow m\mathbb{Z}$ then, for arbitrarily large regions $R$, the system eigenstates $\mid n\rangle$ obey
\begin{align*}
\left\langle U_{\text{f},R}^{p}\right\rangle _{n} & \sim\begin{cases}
 e^{-\alpha\text{vol}\left(R\right)} & p\neq0\mod m\\
 e^{-\alpha\left|\partial R\right|} & p=0\mod m
\end{cases}
\end{align*}
where $p\in \mathbb{Z}$.

\end{appendix}

\bibliography{global}

\begin{thebibliography}{47}%
\makeatletter
\providecommand \@ifxundefined [1]{%
 \@ifx{#1\undefined}
}%
\providecommand \@ifnum [1]{%
 \ifnum #1\expandafter \@firstoftwo
 \else \expandafter \@secondoftwo
 \fi
}%
\providecommand \@ifx [1]{%
 \ifx #1\expandafter \@firstoftwo
 \else \expandafter \@secondoftwo
 \fi
}%
\providecommand \natexlab [1]{#1}%
\providecommand \enquote  [1]{``#1''}%
\providecommand \bibnamefont  [1]{#1}%
\providecommand \bibfnamefont [1]{#1}%
\providecommand \citenamefont [1]{#1}%
\providecommand \href@noop [0]{\@secondoftwo}%
\providecommand \href [0]{\begingroup \@sanitize@url \@href}%
\providecommand \@href[1]{\@@startlink{#1}\@@href}%
\providecommand \@@href[1]{\endgroup#1\@@endlink}%
\providecommand \@sanitize@url [0]{\catcode `\\12\catcode `\$12\catcode
  `\&12\catcode `\#12\catcode `\^12\catcode `\_12\catcode `\%12\relax}%
\providecommand \@@startlink[1]{}%
\providecommand \@@endlink[0]{}%
\providecommand \url  [0]{\begingroup\@sanitize@url \@url }%
\providecommand \@url [1]{\endgroup\@href {#1}{\urlprefix }}%
\providecommand \urlprefix  [0]{URL }%
\providecommand \Eprint [0]{\href }%
\providecommand \doibase [0]{http://dx.doi.org/}%
\providecommand \selectlanguage [0]{\@gobble}%
\providecommand \bibinfo  [0]{\@secondoftwo}%
\providecommand \bibfield  [0]{\@secondoftwo}%
\providecommand \translation [1]{[#1]}%
\providecommand \BibitemOpen [0]{}%
\providecommand \bibitemStop [0]{}%
\providecommand \bibitemNoStop [0]{.\EOS\space}%
\providecommand \EOS [0]{\spacefactor3000\relax}%
\providecommand \BibitemShut  [1]{\csname bibitem#1\endcsname}%
\let\auto@bib@innerbib\@empty
\bibitem [{\citenamefont {Wilczek}(2012)}]{Wilczek12}%
  \BibitemOpen
  \bibfield  {author} {\bibinfo {author} {\bibfnamefont {F.}~\bibnamefont
  {Wilczek}},\ }\href {\doibase 10.1103/PhysRevLett.109.160401} {\bibfield
  {journal} {\bibinfo  {journal} {Phys. Rev. Lett.}\ }\textbf {\bibinfo
  {volume} {109}},\ \bibinfo {pages} {160401} (\bibinfo {year}
  {2012})}\BibitemShut {NoStop}%
\bibitem [{\citenamefont {Watanabe}\ and\ \citenamefont
  {Oshikawa}(2015)}]{Oshikawa15}%
  \BibitemOpen
  \bibfield  {author} {\bibinfo {author} {\bibfnamefont {H.}~\bibnamefont
  {Watanabe}}\ and\ \bibinfo {author} {\bibfnamefont {M.}~\bibnamefont
  {Oshikawa}},\ }\href {\doibase 10.1103/PhysRevLett.114.251603} {\bibfield
  {journal} {\bibinfo  {journal} {Phys. Rev. Lett.}\ }\textbf {\bibinfo
  {volume} {114}},\ \bibinfo {pages} {251603} (\bibinfo {year}
  {2015})}\BibitemShut {NoStop}%
\bibitem [{\citenamefont {{Khemani}}\ \emph {et~al.}(2016)\citenamefont
  {{Khemani}}, \citenamefont {{Lazarides}}, \citenamefont {{Moessner}},\ and\
  \citenamefont {{Sondhi}}}]{Khemani15}%
  \BibitemOpen
  \bibfield  {author} {\bibinfo {author} {\bibfnamefont {V.}~\bibnamefont
  {{Khemani}}}, \bibinfo {author} {\bibfnamefont {A.}~\bibnamefont
  {{Lazarides}}}, \bibinfo {author} {\bibfnamefont {R.}~\bibnamefont
  {{Moessner}}}, \ and\ \bibinfo {author} {\bibfnamefont {S.~L.}\ \bibnamefont
  {{Sondhi}}},\ }\href {\doibase 10.1103/PhysRevLett.116.250401} {\bibfield
  {journal} {\bibinfo  {journal} {Physical Review Letters}\ }\textbf {\bibinfo
  {volume} {116}},\ \bibinfo {eid} {250401} (\bibinfo {year} {2016})},\ \Eprint
  {http://arxiv.org/abs/1508.03344} {arXiv:1508.03344 [cond-mat.dis-nn]}
  \BibitemShut {NoStop}%
\bibitem [{\citenamefont {Lazarides}\ \emph {et~al.}(2015)\citenamefont
  {Lazarides}, \citenamefont {Das},\ and\ \citenamefont
  {Moessner}}]{Lazarides14}%
  \BibitemOpen
  \bibfield  {author} {\bibinfo {author} {\bibfnamefont {A.}~\bibnamefont
  {Lazarides}}, \bibinfo {author} {\bibfnamefont {A.}~\bibnamefont {Das}}, \
  and\ \bibinfo {author} {\bibfnamefont {R.}~\bibnamefont {Moessner}},\ }\href
  {\doibase 10.1103/PhysRevLett.115.030402} {\bibfield  {journal} {\bibinfo
  {journal} {Phys. Rev. Lett.}\ }\textbf {\bibinfo {volume} {115}},\ \bibinfo
  {pages} {030402} (\bibinfo {year} {2015})}\BibitemShut {NoStop}%
\bibitem [{\citenamefont {Ponte}\ \emph
  {et~al.}(2015{\natexlab{a}})\citenamefont {Ponte}, \citenamefont
  {Papi\ifmmode~\acute{c}\else \'{c}\fi{}}, \citenamefont {Huveneers},\ and\
  \citenamefont {Abanin}}]{Ponte15}%
  \BibitemOpen
  \bibfield  {author} {\bibinfo {author} {\bibfnamefont {P.}~\bibnamefont
  {Ponte}}, \bibinfo {author} {\bibfnamefont {Z.}~\bibnamefont
  {Papi\ifmmode~\acute{c}\else \'{c}\fi{}}}, \bibinfo {author} {\bibfnamefont
  {F.}~\bibnamefont {Huveneers}}, \ and\ \bibinfo {author} {\bibfnamefont
  {D.~A.}\ \bibnamefont {Abanin}},\ }\href {\doibase
  10.1103/PhysRevLett.114.140401} {\bibfield  {journal} {\bibinfo  {journal}
  {Phys. Rev. Lett.}\ }\textbf {\bibinfo {volume} {114}},\ \bibinfo {pages}
  {140401} (\bibinfo {year} {2015}{\natexlab{a}})}\BibitemShut {NoStop}%
\bibitem [{\citenamefont {Ponte}\ \emph
  {et~al.}(2015{\natexlab{b}})\citenamefont {Ponte}, \citenamefont {Chandran},
  \citenamefont {Papić},\ and\ \citenamefont {Abanin}}]{Ponte15b}%
  \BibitemOpen
  \bibfield  {author} {\bibinfo {author} {\bibfnamefont {P.}~\bibnamefont
  {Ponte}}, \bibinfo {author} {\bibfnamefont {A.}~\bibnamefont {Chandran}},
  \bibinfo {author} {\bibfnamefont {Z.}~\bibnamefont {Papić}}, \ and\ \bibinfo
  {author} {\bibfnamefont {D.~A.}\ \bibnamefont {Abanin}},\ }\href {\doibase
  http://dx.doi.org/10.1016/j.aop.2014.11.008} {\bibfield  {journal} {\bibinfo
  {journal} {Annals of Physics}\ }\textbf {\bibinfo {volume} {353}},\ \bibinfo
  {pages} {196 } (\bibinfo {year} {2015}{\natexlab{b}})}\BibitemShut {NoStop}%
\bibitem [{\citenamefont {Abanin}\ \emph {et~al.}(2014)\citenamefont {Abanin},
  \citenamefont {De~Roeck},\ and\ \citenamefont {Huveneers}}]{Abanin14}%
  \BibitemOpen
  \bibfield  {author} {\bibinfo {author} {\bibfnamefont {D.}~\bibnamefont
  {Abanin}}, \bibinfo {author} {\bibfnamefont {W.}~\bibnamefont {De~Roeck}}, \
  and\ \bibinfo {author} {\bibfnamefont {F.}~\bibnamefont {Huveneers}},\
  }\href@noop {} {\bibfield  {journal} {\bibinfo  {journal} {arXiv preprint
  arXiv:1412.4752}\ } (\bibinfo {year} {2014})}\BibitemShut {NoStop}%
\bibitem [{\citenamefont {D'Alessio}\ and\ \citenamefont
  {Rigol}(2014)}]{Rigol14}%
  \BibitemOpen
  \bibfield  {author} {\bibinfo {author} {\bibfnamefont {L.}~\bibnamefont
  {D'Alessio}}\ and\ \bibinfo {author} {\bibfnamefont {M.}~\bibnamefont
  {Rigol}},\ }\href {\doibase 10.1103/PhysRevX.4.041048} {\bibfield  {journal}
  {\bibinfo  {journal} {Phys. Rev. X}\ }\textbf {\bibinfo {volume} {4}},\
  \bibinfo {pages} {041048} (\bibinfo {year} {2014})}\BibitemShut {NoStop}%
\bibitem [{\citenamefont {Huse}\ \emph {et~al.}(2013)\citenamefont {Huse},
  \citenamefont {Nandkishore}, \citenamefont {Oganesyan}, \citenamefont {Pal},\
  and\ \citenamefont {Sondhi}}]{Huse13}%
  \BibitemOpen
  \bibfield  {author} {\bibinfo {author} {\bibfnamefont {D.~A.}\ \bibnamefont
  {Huse}}, \bibinfo {author} {\bibfnamefont {R.}~\bibnamefont {Nandkishore}},
  \bibinfo {author} {\bibfnamefont {V.}~\bibnamefont {Oganesyan}}, \bibinfo
  {author} {\bibfnamefont {A.}~\bibnamefont {Pal}}, \ and\ \bibinfo {author}
  {\bibfnamefont {S.~L.}\ \bibnamefont {Sondhi}},\ }\href
  {http://link.aps.org/doi/10.1103/PhysRevB.88.014206} {\bibfield  {journal}
  {\bibinfo  {journal} {Phys. Rev. B}\ }\textbf {\bibinfo {volume} {88}},\
  \bibinfo {pages} {014206} (\bibinfo {year} {2013})}\BibitemShut {NoStop}%
\bibitem [{\citenamefont {Pekker}\ \emph {et~al.}(2014)\citenamefont {Pekker},
  \citenamefont {Refael}, \citenamefont {Altman}, \citenamefont {Demler},\ and\
  \citenamefont {Oganesyan}}]{PekkerHilbertGlass}%
  \BibitemOpen
  \bibfield  {author} {\bibinfo {author} {\bibfnamefont {D.}~\bibnamefont
  {Pekker}}, \bibinfo {author} {\bibfnamefont {G.}~\bibnamefont {Refael}},
  \bibinfo {author} {\bibfnamefont {E.}~\bibnamefont {Altman}}, \bibinfo
  {author} {\bibfnamefont {E.}~\bibnamefont {Demler}}, \ and\ \bibinfo {author}
  {\bibfnamefont {V.}~\bibnamefont {Oganesyan}},\ }\href
  {http://link.aps.org/doi/10.1103/PhysRevX.4.011052} {\bibfield  {journal}
  {\bibinfo  {journal} {Phys. Rev. X}\ }\textbf {\bibinfo {volume} {4}},\
  \bibinfo {pages} {011052} (\bibinfo {year} {2014})}\BibitemShut {NoStop}%
\bibitem [{\citenamefont {Bauer}\ and\ \citenamefont {Nayak}(2013)}]{Bauer13}%
  \BibitemOpen
  \bibfield  {author} {\bibinfo {author} {\bibfnamefont {B.}~\bibnamefont
  {Bauer}}\ and\ \bibinfo {author} {\bibfnamefont {C.}~\bibnamefont {Nayak}},\
  }\href {http://stacks.iop.org/1742-5468/2013/i=09/a=P09005} {\bibfield
  {journal} {\bibinfo  {journal} {Journal of Statistical Mechanics: Theory and
  Experiment}\ }\textbf {\bibinfo {volume} {2013}},\ \bibinfo {pages} {P09005}
  (\bibinfo {year} {2013})}\BibitemShut {NoStop}%
\bibitem [{\citenamefont {Chandran}\ \emph {et~al.}(2014)\citenamefont
  {Chandran}, \citenamefont {Khemani}, \citenamefont {Laumann},\ and\
  \citenamefont {Sondhi}}]{Chandran14}%
  \BibitemOpen
  \bibfield  {author} {\bibinfo {author} {\bibfnamefont {A.}~\bibnamefont
  {Chandran}}, \bibinfo {author} {\bibfnamefont {V.}~\bibnamefont {Khemani}},
  \bibinfo {author} {\bibfnamefont {C.~R.}\ \bibnamefont {Laumann}}, \ and\
  \bibinfo {author} {\bibfnamefont {S.~L.}\ \bibnamefont {Sondhi}},\ }\href
  {\doibase 10.1103/PhysRevB.89.144201} {\bibfield  {journal} {\bibinfo
  {journal} {Phys. Rev. B}\ }\textbf {\bibinfo {volume} {89}},\ \bibinfo
  {pages} {144201} (\bibinfo {year} {2014})}\BibitemShut {NoStop}%
\bibitem [{\citenamefont {Bahri}\ \emph {et~al.}(2015)\citenamefont {Bahri},
  \citenamefont {Vosk}, \citenamefont {Altman},\ and\ \citenamefont
  {Vishwanath}}]{Bahri15}%
  \BibitemOpen
  \bibfield  {author} {\bibinfo {author} {\bibfnamefont {Y.}~\bibnamefont
  {Bahri}}, \bibinfo {author} {\bibfnamefont {R.}~\bibnamefont {Vosk}},
  \bibinfo {author} {\bibfnamefont {E.}~\bibnamefont {Altman}}, \ and\ \bibinfo
  {author} {\bibfnamefont {A.}~\bibnamefont {Vishwanath}},\ }\href
  {http://dx.doi.org/10.1038/ncomms8341} {\bibfield  {journal} {\bibinfo
  {journal} {Nat Commun}\ }\textbf {\bibinfo {volume} {6}} (\bibinfo {year}
  {2015})}\BibitemShut {NoStop}%
\bibitem [{\citenamefont {Potter}\ and\ \citenamefont
  {Vishwanath}(2015)}]{Potter15}%
  \BibitemOpen
  \bibfield  {author} {\bibinfo {author} {\bibfnamefont {A.~C.}\ \bibnamefont
  {Potter}}\ and\ \bibinfo {author} {\bibfnamefont {A.}~\bibnamefont
  {Vishwanath}},\ }\href@noop {} {\bibfield  {journal} {\bibinfo  {journal}
  {arXiv preprint arXiv:1506.00592}\ } (\bibinfo {year} {2015})}\BibitemShut
  {NoStop}%
\bibitem [{\citenamefont {Anderson}(1958)}]{Anderson58}%
  \BibitemOpen
  \bibfield  {author} {\bibinfo {author} {\bibfnamefont {P.~W.}\ \bibnamefont
  {Anderson}},\ }\href {\doibase 10.1103/PhysRev.109.1492} {\bibfield
  {journal} {\bibinfo  {journal} {Phys. Rev.}\ }\textbf {\bibinfo {volume}
  {109}},\ \bibinfo {pages} {1492} (\bibinfo {year} {1958})}\BibitemShut
  {NoStop}%
\bibitem [{\citenamefont {Basko}\ \emph {et~al.}(2006)\citenamefont {Basko},
  \citenamefont {Aleiner},\ and\ \citenamefont {Altshuler}}]{Basko06}%
  \BibitemOpen
  \bibfield  {author} {\bibinfo {author} {\bibfnamefont {D.~M.}\ \bibnamefont
  {Basko}}, \bibinfo {author} {\bibfnamefont {I.~L.}\ \bibnamefont {Aleiner}},
  \ and\ \bibinfo {author} {\bibfnamefont {B.~L.}\ \bibnamefont {Altshuler}},\
  }\href {\doibase 10.1016/j.aop.2005.11.014} {\bibfield  {journal} {\bibinfo
  {journal} {Annals of Physics}\ }\textbf {\bibinfo {volume} {321}},\ \bibinfo
  {pages} {1126} (\bibinfo {year} {2006})}\BibitemShut {NoStop}%
\bibitem [{\citenamefont {Pal}\ and\ \citenamefont {Huse}(2010)}]{PalHuse}%
  \BibitemOpen
  \bibfield  {author} {\bibinfo {author} {\bibfnamefont {A.}~\bibnamefont
  {Pal}}\ and\ \bibinfo {author} {\bibfnamefont {D.~A.}\ \bibnamefont {Huse}},\
  }\href {\doibase 10.1103/PhysRevB.82.174411} {\bibfield  {journal} {\bibinfo
  {journal} {Phys. Rev. B}\ }\textbf {\bibinfo {volume} {82}},\ \bibinfo
  {pages} {174411} (\bibinfo {year} {2010})}\BibitemShut {NoStop}%
\bibitem [{\citenamefont {Oganesyan}\ and\ \citenamefont
  {Huse}(2007)}]{OganesyanHuse}%
  \BibitemOpen
  \bibfield  {author} {\bibinfo {author} {\bibfnamefont {V.}~\bibnamefont
  {Oganesyan}}\ and\ \bibinfo {author} {\bibfnamefont {D.~A.}\ \bibnamefont
  {Huse}},\ }\href {\doibase 10.1103/PhysRevB.75.155111} {\bibfield  {journal}
  {\bibinfo  {journal} {Phys. Rev. B}\ }\textbf {\bibinfo {volume} {75}},\
  \bibinfo {pages} {155111} (\bibinfo {year} {2007})}\BibitemShut {NoStop}%
\bibitem [{\citenamefont {Nandkishore}\ and\ \citenamefont
  {Huse}(2015)}]{Nandkishore14}%
  \BibitemOpen
  \bibfield  {author} {\bibinfo {author} {\bibfnamefont {R.}~\bibnamefont
  {Nandkishore}}\ and\ \bibinfo {author} {\bibfnamefont {D.~A.}\ \bibnamefont
  {Huse}},\ }\href {\doibase 10.1146/annurev-conmatphys-031214-014726}
  {\bibfield  {journal} {\bibinfo  {journal} {Annual Review of Condensed Matter
  Physics}\ }\textbf {\bibinfo {volume} {6}},\ \bibinfo {pages} {15} (\bibinfo
  {year} {2015})}\BibitemShut {NoStop}%
\bibitem [{\citenamefont {Altman}\ and\ \citenamefont
  {Vosk}(2015)}]{AltmanVosk}%
  \BibitemOpen
  \bibfield  {author} {\bibinfo {author} {\bibfnamefont {E.}~\bibnamefont
  {Altman}}\ and\ \bibinfo {author} {\bibfnamefont {R.}~\bibnamefont {Vosk}},\
  }\href {\doibase 10.1146/annurev-conmatphys-031214-014701} {\bibfield
  {journal} {\bibinfo  {journal} {Annual Review of Condensed Matter Physics}\
  }\textbf {\bibinfo {volume} {6}},\ \bibinfo {pages} {383} (\bibinfo {year}
  {2015})}\BibitemShut {NoStop}%
\bibitem [{\citenamefont {Else}\ \emph {et~al.}(2016)\citenamefont {Else},
  \citenamefont {Bauer},\ and\ \citenamefont {Nayak}}]{Q}%
  \BibitemOpen
  \bibfield  {author} {\bibinfo {author} {\bibfnamefont {D.~V.}\ \bibnamefont
  {Else}}, \bibinfo {author} {\bibfnamefont {B.}~\bibnamefont {Bauer}}, \ and\
  \bibinfo {author} {\bibfnamefont {C.}~\bibnamefont {Nayak}},\ }\href
  {\doibase 10.1103/PhysRevLett.117.090402} {\bibfield  {journal} {\bibinfo
  {journal} {Phys. Rev. Lett.}\ }\textbf {\bibinfo {volume} {117}},\ \bibinfo
  {pages} {090402} (\bibinfo {year} {2016})}\BibitemShut {NoStop}%
\bibitem [{\citenamefont {{Yao}}\ \emph {et~al.}(2016)\citenamefont {{Yao}},
  \citenamefont {{Potter}}, \citenamefont {{Potirniche}},\ and\ \citenamefont
  {{Vishwanath}}}]{YaoDTC}%
  \BibitemOpen
  \bibfield  {author} {\bibinfo {author} {\bibfnamefont {N.~Y.}\ \bibnamefont
  {{Yao}}}, \bibinfo {author} {\bibfnamefont {A.~C.}\ \bibnamefont {{Potter}}},
  \bibinfo {author} {\bibfnamefont {I.-D.}\ \bibnamefont {{Potirniche}}}, \
  and\ \bibinfo {author} {\bibfnamefont {A.}~\bibnamefont {{Vishwanath}}},\
  }\href@noop {} {\bibfield  {journal} {\bibinfo  {journal} {ArXiv e-prints}\ }
  (\bibinfo {year} {2016})},\ \Eprint {http://arxiv.org/abs/1608.02589}
  {arXiv:1608.02589 [cond-mat.dis-nn]} \BibitemShut {NoStop}%
\bibitem [{\citenamefont {von Keyserlingk}\ and\ \citenamefont
  {Sondhi}(2016)}]{vonkeyserlingk16b}%
  \BibitemOpen
  \bibfield  {author} {\bibinfo {author} {\bibfnamefont {C.~W.}\ \bibnamefont
  {von Keyserlingk}}\ and\ \bibinfo {author} {\bibfnamefont {S.~L.}\
  \bibnamefont {Sondhi}},\ }\href {\doibase 10.1103/PhysRevB.93.245146}
  {\bibfield  {journal} {\bibinfo  {journal} {Phys. Rev. B}\ }\textbf {\bibinfo
  {volume} {93}},\ \bibinfo {pages} {245146} (\bibinfo {year}
  {2016})}\BibitemShut {NoStop}%
\bibitem [{\citenamefont {von Keyserlingk}\ \emph {et~al.}(2016)\citenamefont
  {von Keyserlingk}, \citenamefont {Khemani},\ and\ \citenamefont
  {Sondhi}}]{CVS}%
  \BibitemOpen
  \bibfield  {author} {\bibinfo {author} {\bibfnamefont {C.~W.}\ \bibnamefont
  {von Keyserlingk}}, \bibinfo {author} {\bibfnamefont {V.}~\bibnamefont
  {Khemani}}, \ and\ \bibinfo {author} {\bibfnamefont {S.~L.}\ \bibnamefont
  {Sondhi}},\ }\href {\doibase 10.1103/PhysRevB.94.085112} {\bibfield
  {journal} {\bibinfo  {journal} {Phys. Rev. B}\ }\textbf {\bibinfo {volume}
  {94}},\ \bibinfo {pages} {085112} (\bibinfo {year} {2016})}\BibitemShut
  {NoStop}%
\bibitem [{\citenamefont {{Else}}\ \emph {et~al.}(2016)\citenamefont {{Else}},
  \citenamefont {{Bauer}},\ and\ \citenamefont {{Nayak}}}]{Q2}%
  \BibitemOpen
  \bibfield  {author} {\bibinfo {author} {\bibfnamefont {D.~V.}\ \bibnamefont
  {{Else}}}, \bibinfo {author} {\bibfnamefont {B.}~\bibnamefont {{Bauer}}}, \
  and\ \bibinfo {author} {\bibfnamefont {C.}~\bibnamefont {{Nayak}}},\
  }\href@noop {} {\bibfield  {journal} {\bibinfo  {journal} {ArXiv e-prints}\ }
  (\bibinfo {year} {2016})},\ \Eprint {http://arxiv.org/abs/1607.05277}
  {arXiv:1607.05277 [cond-mat.stat-mech]} \BibitemShut {NoStop}%
\bibitem [{\citenamefont {{Zhang}}\ \emph {et~al.}(2016)\citenamefont
  {{Zhang}}, \citenamefont {{Hess}}, \citenamefont {{Kyprianidis}},
  \citenamefont {{Becker}}, \citenamefont {{Lee}}, \citenamefont {{Smith}},
  \citenamefont {{Pagano}}, \citenamefont {{Potirniche}}, \citenamefont
  {{Potter}}, \citenamefont {{Vishwanath}}, \citenamefont {{Yao}},\ and\
  \citenamefont {{Monroe}}}]{MonroeTCExp}%
  \BibitemOpen
  \bibfield  {author} {\bibinfo {author} {\bibfnamefont {J.}~\bibnamefont
  {{Zhang}}}, \bibinfo {author} {\bibfnamefont {P.~W.}\ \bibnamefont {{Hess}}},
  \bibinfo {author} {\bibfnamefont {A.}~\bibnamefont {{Kyprianidis}}}, \bibinfo
  {author} {\bibfnamefont {P.}~\bibnamefont {{Becker}}}, \bibinfo {author}
  {\bibfnamefont {A.}~\bibnamefont {{Lee}}}, \bibinfo {author} {\bibfnamefont
  {J.}~\bibnamefont {{Smith}}}, \bibinfo {author} {\bibfnamefont
  {G.}~\bibnamefont {{Pagano}}}, \bibinfo {author} {\bibfnamefont {I.-D.}\
  \bibnamefont {{Potirniche}}}, \bibinfo {author} {\bibfnamefont {A.~C.}\
  \bibnamefont {{Potter}}}, \bibinfo {author} {\bibfnamefont {A.}~\bibnamefont
  {{Vishwanath}}}, \bibinfo {author} {\bibfnamefont {N.~Y.}\ \bibnamefont
  {{Yao}}}, \ and\ \bibinfo {author} {\bibfnamefont {C.}~\bibnamefont
  {{Monroe}}},\ }\href@noop {} {\bibfield  {journal} {\bibinfo  {journal}
  {ArXiv e-prints}\ } (\bibinfo {year} {2016})},\ \Eprint
  {http://arxiv.org/abs/1609.08684} {arXiv:1609.08684 [quant-ph]} \BibitemShut
  {NoStop}%
\bibitem [{\citenamefont {{Choi}}\ \emph {et~al.}(2016)\citenamefont {{Choi}},
  \citenamefont {{Choi}}, \citenamefont {{Landig}}, \citenamefont {{Kucsko}},
  \citenamefont {{Zhou}}, \citenamefont {{Isoya}}, \citenamefont {{Jelezko}},
  \citenamefont {{Onoda}}, \citenamefont {{Sumiya}}, \citenamefont {{Khemani}},
  \citenamefont {{von Keyserlingk}}, \citenamefont {{Yao}}, \citenamefont
  {{Demler}},\ and\ \citenamefont {{Lukin}}}]{MishaTCExp}%
  \BibitemOpen
  \bibfield  {author} {\bibinfo {author} {\bibfnamefont {S.}~\bibnamefont
  {{Choi}}}, \bibinfo {author} {\bibfnamefont {J.}~\bibnamefont {{Choi}}},
  \bibinfo {author} {\bibfnamefont {R.}~\bibnamefont {{Landig}}}, \bibinfo
  {author} {\bibfnamefont {G.}~\bibnamefont {{Kucsko}}}, \bibinfo {author}
  {\bibfnamefont {H.}~\bibnamefont {{Zhou}}}, \bibinfo {author} {\bibfnamefont
  {J.}~\bibnamefont {{Isoya}}}, \bibinfo {author} {\bibfnamefont
  {F.}~\bibnamefont {{Jelezko}}}, \bibinfo {author} {\bibfnamefont
  {S.}~\bibnamefont {{Onoda}}}, \bibinfo {author} {\bibfnamefont
  {H.}~\bibnamefont {{Sumiya}}}, \bibinfo {author} {\bibfnamefont
  {V.}~\bibnamefont {{Khemani}}}, \bibinfo {author} {\bibfnamefont
  {C.}~\bibnamefont {{von Keyserlingk}}}, \bibinfo {author} {\bibfnamefont
  {N.~Y.}\ \bibnamefont {{Yao}}}, \bibinfo {author} {\bibfnamefont
  {E.}~\bibnamefont {{Demler}}}, \ and\ \bibinfo {author} {\bibfnamefont
  {M.~D.}\ \bibnamefont {{Lukin}}},\ }\href@noop {} {\bibfield  {journal}
  {\bibinfo  {journal} {ArXiv e-prints}\ } (\bibinfo {year} {2016})},\ \Eprint
  {http://arxiv.org/abs/1610.08057} {arXiv:1610.08057 [quant-ph]} \BibitemShut
  {NoStop}%
\bibitem [{Note1()}]{Note1}%
  \BibitemOpen
  \bibinfo {note} {We restrict to unitary symmetries for this
  discussion.}\BibitemShut {Stop}%
\bibitem [{\citenamefont {{Po}}\ \emph {et~al.}(2016)\citenamefont {{Po}},
  \citenamefont {{Fidkowski}}, \citenamefont {{Morimoto}}, \citenamefont
  {{Potter}},\ and\ \citenamefont {{Vishwanath}}}]{Po2016}%
  \BibitemOpen
  \bibfield  {author} {\bibinfo {author} {\bibfnamefont {H.~C.}\ \bibnamefont
  {{Po}}}, \bibinfo {author} {\bibfnamefont {L.}~\bibnamefont {{Fidkowski}}},
  \bibinfo {author} {\bibfnamefont {T.}~\bibnamefont {{Morimoto}}}, \bibinfo
  {author} {\bibfnamefont {A.~C.}\ \bibnamefont {{Potter}}}, \ and\ \bibinfo
  {author} {\bibfnamefont {A.}~\bibnamefont {{Vishwanath}}},\ }\href@noop {}
  {\bibfield  {journal} {\bibinfo  {journal} {ArXiv e-prints}\ } (\bibinfo
  {year} {2016})},\ \Eprint {http://arxiv.org/abs/1609.00006} {arXiv:1609.00006
  [cond-mat.dis-nn]} \BibitemShut {NoStop}%
\bibitem [{Note2()}]{Note2}%
  \BibitemOpen
  \bibinfo {note} {Thus, perhaps counterintuitively, time translation symmetry
  is more similar to the case of a global symmetry rather than spatial
  translation symmetry.}\BibitemShut {Stop}%
\bibitem [{Note3()}]{Note3}%
  \BibitemOpen
  \bibinfo {note} {See e.g. Ref.~\protect \rev@citealpnum
  {PaisRMP}.}\BibitemShut {Stop}%
\bibitem [{\citenamefont {Lieb}\ and\ \citenamefont {Robinson}()}]{Lieb72}%
  \BibitemOpen
  \bibfield  {author} {\bibinfo {author} {\bibfnamefont {E.~H.}\ \bibnamefont
  {Lieb}}\ and\ \bibinfo {author} {\bibfnamefont {D.~W.}\ \bibnamefont
  {Robinson}},\ }\href {\doibase 10.1007/BF01645779} {\bibfield  {journal}
  {\bibinfo  {journal} {Communications in Mathematical Physics}\ }\textbf
  {\bibinfo {volume} {28}},\ \bibinfo {pages} {251}}\BibitemShut {NoStop}%
\bibitem [{\citenamefont {Agarwal}\ \emph {et~al.}(2015)\citenamefont
  {Agarwal}, \citenamefont {Gopalakrishnan}, \citenamefont {Knap},
  \citenamefont {M\"uller},\ and\ \citenamefont {Demler}}]{AgarwalGriffiths}%
  \BibitemOpen
  \bibfield  {author} {\bibinfo {author} {\bibfnamefont {K.}~\bibnamefont
  {Agarwal}}, \bibinfo {author} {\bibfnamefont {S.}~\bibnamefont
  {Gopalakrishnan}}, \bibinfo {author} {\bibfnamefont {M.}~\bibnamefont
  {Knap}}, \bibinfo {author} {\bibfnamefont {M.}~\bibnamefont {M\"uller}}, \
  and\ \bibinfo {author} {\bibfnamefont {E.}~\bibnamefont {Demler}},\ }\href
  {\doibase 10.1103/PhysRevLett.114.160401} {\bibfield  {journal} {\bibinfo
  {journal} {Phys. Rev. Lett.}\ }\textbf {\bibinfo {volume} {114}},\ \bibinfo
  {pages} {160401} (\bibinfo {year} {2015})}\BibitemShut {NoStop}%
\bibitem [{\citenamefont {Lazarides}\ \emph {et~al.}(2014)\citenamefont
  {Lazarides}, \citenamefont {Das},\ and\ \citenamefont
  {Moessner}}]{Lazarides14PRE}%
  \BibitemOpen
  \bibfield  {author} {\bibinfo {author} {\bibfnamefont {A.}~\bibnamefont
  {Lazarides}}, \bibinfo {author} {\bibfnamefont {A.}~\bibnamefont {Das}}, \
  and\ \bibinfo {author} {\bibfnamefont {R.}~\bibnamefont {Moessner}},\ }\href
  {\doibase 10.1103/PhysRevE.90.012110} {\bibfield  {journal} {\bibinfo
  {journal} {Phys. Rev. E}\ }\textbf {\bibinfo {volume} {90}},\ \bibinfo
  {pages} {012110} (\bibinfo {year} {2014})}\BibitemShut {NoStop}%
\bibitem [{\citenamefont {Serbyn}\ \emph
  {et~al.}(2013{\natexlab{a}})\citenamefont {Serbyn}, \citenamefont
  {Papi\ifmmode~\acute{c}\else \'{c}\fi{}},\ and\ \citenamefont
  {Abanin}}]{Serbyn13a}%
  \BibitemOpen
  \bibfield  {author} {\bibinfo {author} {\bibfnamefont {M.}~\bibnamefont
  {Serbyn}}, \bibinfo {author} {\bibfnamefont {Z.}~\bibnamefont
  {Papi\ifmmode~\acute{c}\else \'{c}\fi{}}}, \ and\ \bibinfo {author}
  {\bibfnamefont {D.~A.}\ \bibnamefont {Abanin}},\ }\href {\doibase
  10.1103/PhysRevLett.110.260601} {\bibfield  {journal} {\bibinfo  {journal}
  {Phys. Rev. Lett.}\ }\textbf {\bibinfo {volume} {110}},\ \bibinfo {pages}
  {260601} (\bibinfo {year} {2013}{\natexlab{a}})}\BibitemShut {NoStop}%
\bibitem [{\citenamefont {Serbyn}\ \emph
  {et~al.}(2013{\natexlab{b}})\citenamefont {Serbyn}, \citenamefont
  {Papi\ifmmode~\acute{c}\else \'{c}\fi{}},\ and\ \citenamefont
  {Abanin}}]{Serbyn13cons}%
  \BibitemOpen
  \bibfield  {author} {\bibinfo {author} {\bibfnamefont {M.}~\bibnamefont
  {Serbyn}}, \bibinfo {author} {\bibfnamefont {Z.}~\bibnamefont
  {Papi\ifmmode~\acute{c}\else \'{c}\fi{}}}, \ and\ \bibinfo {author}
  {\bibfnamefont {D.~A.}\ \bibnamefont {Abanin}},\ }\href
  {http://link.aps.org/doi/10.1103/PhysRevLett.111.127201} {\bibfield
  {journal} {\bibinfo  {journal} {Phys. Rev. Lett.}\ }\textbf {\bibinfo
  {volume} {111}},\ \bibinfo {pages} {127201} (\bibinfo {year}
  {2013}{\natexlab{b}})}\BibitemShut {NoStop}%
\bibitem [{\citenamefont {Huse}\ \emph {et~al.}(2014)\citenamefont {Huse},
  \citenamefont {Nandkishore},\ and\ \citenamefont {Oganesyan}}]{Huse14}%
  \BibitemOpen
  \bibfield  {author} {\bibinfo {author} {\bibfnamefont {D.~A.}\ \bibnamefont
  {Huse}}, \bibinfo {author} {\bibfnamefont {R.}~\bibnamefont {Nandkishore}}, \
  and\ \bibinfo {author} {\bibfnamefont {V.}~\bibnamefont {Oganesyan}},\ }\href
  {\doibase 10.1103/PhysRevB.90.174202} {\bibfield  {journal} {\bibinfo
  {journal} {Phys. Rev. B}\ }\textbf {\bibinfo {volume} {90}},\ \bibinfo
  {pages} {174202} (\bibinfo {year} {2014})}\BibitemShut {NoStop}%
\bibitem [{\citenamefont {{Ros}}\ \emph {et~al.}(2015)\citenamefont {{Ros}},
  \citenamefont {{M{\"u}ller}},\ and\ \citenamefont {{Scardicchio}}}]{Ros15}%
  \BibitemOpen
  \bibfield  {author} {\bibinfo {author} {\bibfnamefont {V.}~\bibnamefont
  {{Ros}}}, \bibinfo {author} {\bibfnamefont {M.}~\bibnamefont {{M{\"u}ller}}},
  \ and\ \bibinfo {author} {\bibfnamefont {A.}~\bibnamefont {{Scardicchio}}},\
  }\href {\doibase 10.1016/j.nuclphysb.2014.12.014} {\bibfield  {journal}
  {\bibinfo  {journal} {Nuclear Physics B}\ }\textbf {\bibinfo {volume}
  {891}},\ \bibinfo {pages} {420} (\bibinfo {year} {2015})}\BibitemShut
  {NoStop}%
\bibitem [{\citenamefont {Chandran}\ \emph {et~al.}(2015)\citenamefont
  {Chandran}, \citenamefont {Kim}, \citenamefont {Vidal},\ and\ \citenamefont
  {Abanin}}]{Chandran15b}%
  \BibitemOpen
  \bibfield  {author} {\bibinfo {author} {\bibfnamefont {A.}~\bibnamefont
  {Chandran}}, \bibinfo {author} {\bibfnamefont {I.~H.}\ \bibnamefont {Kim}},
  \bibinfo {author} {\bibfnamefont {G.}~\bibnamefont {Vidal}}, \ and\ \bibinfo
  {author} {\bibfnamefont {D.~A.}\ \bibnamefont {Abanin}},\ }\href
  {http://link.aps.org/doi/10.1103/PhysRevB.91.085425} {\bibfield  {journal}
  {\bibinfo  {journal} {Phys. Rev. B}\ }\textbf {\bibinfo {volume} {91}},\
  \bibinfo {pages} {085425} (\bibinfo {year} {2015})}\BibitemShut {NoStop}%
\bibitem [{\citenamefont {{Swingle}}\ and\ \citenamefont
  {{Chowdhury}}(2016)}]{Swingle16}%
  \BibitemOpen
  \bibfield  {author} {\bibinfo {author} {\bibfnamefont {B.}~\bibnamefont
  {{Swingle}}}\ and\ \bibinfo {author} {\bibfnamefont {D.}~\bibnamefont
  {{Chowdhury}}},\ }\href@noop {} {\bibfield  {journal} {\bibinfo  {journal}
  {ArXiv e-prints}\ } (\bibinfo {year} {2016})},\ \Eprint
  {http://arxiv.org/abs/1608.03280} {arXiv:1608.03280 [cond-mat.str-el]}
  \BibitemShut {NoStop}%
\bibitem [{\citenamefont {{Huang}}\ \emph {et~al.}(2016)\citenamefont
  {{Huang}}, \citenamefont {{Zhang}},\ and\ \citenamefont {{Chen}}}]{Huang16}%
  \BibitemOpen
  \bibfield  {author} {\bibinfo {author} {\bibfnamefont {Y.}~\bibnamefont
  {{Huang}}}, \bibinfo {author} {\bibfnamefont {Y.-L.}\ \bibnamefont
  {{Zhang}}}, \ and\ \bibinfo {author} {\bibfnamefont {X.}~\bibnamefont
  {{Chen}}},\ }\href@noop {} {\bibfield  {journal} {\bibinfo  {journal} {ArXiv
  e-prints}\ } (\bibinfo {year} {2016})},\ \Eprint
  {http://arxiv.org/abs/1608.01091} {arXiv:1608.01091 [cond-mat.dis-nn]}
  \BibitemShut {NoStop}%
\bibitem [{\citenamefont {{Fan}}\ \emph {et~al.}(2016)\citenamefont {{Fan}},
  \citenamefont {{Zhang}}, \citenamefont {{Shen}},\ and\ \citenamefont
  {{Zhai}}}]{Fan16}%
  \BibitemOpen
  \bibfield  {author} {\bibinfo {author} {\bibfnamefont {R.}~\bibnamefont
  {{Fan}}}, \bibinfo {author} {\bibfnamefont {P.}~\bibnamefont {{Zhang}}},
  \bibinfo {author} {\bibfnamefont {H.}~\bibnamefont {{Shen}}}, \ and\ \bibinfo
  {author} {\bibfnamefont {H.}~\bibnamefont {{Zhai}}},\ }\href@noop {}
  {\bibfield  {journal} {\bibinfo  {journal} {ArXiv e-prints}\ } (\bibinfo
  {year} {2016})},\ \Eprint {http://arxiv.org/abs/1608.01914} {arXiv:1608.01914
  [cond-mat.quant-gas]} \BibitemShut {NoStop}%
\bibitem [{\citenamefont {{Chen}}(2016)}]{Chen16}%
  \BibitemOpen
  \bibfield  {author} {\bibinfo {author} {\bibfnamefont {Y.}~\bibnamefont
  {{Chen}}},\ }\href@noop {} {\bibfield  {journal} {\bibinfo  {journal} {ArXiv
  e-prints}\ } (\bibinfo {year} {2016})},\ \Eprint
  {http://arxiv.org/abs/1608.02765} {arXiv:1608.02765 [cond-mat.dis-nn]}
  \BibitemShut {NoStop}%
\bibitem [{Note4()}]{Note4}%
  \BibitemOpen
  \bibinfo {note} {We note that for perturbed drives which preserve Ising
  symmetry ($h^y = h^z = 0, h^x\not =0$), the $\protect \mathbb {Z}_2$ symmetry
  breaking is still diagnosed by the $\sigma _i^z$ operators. On the other
  hand, order parameters for TTSB are defined in terms of dressed l-bit
  operators $\tau _i^z$.}\BibitemShut {Stop}%
\bibitem [{\citenamefont {Gütschow}\ \emph {et~al.}(2010)\citenamefont
  {Gütschow}, \citenamefont {Uphoff}, \citenamefont {Werner},\ and\
  \citenamefont {Zimborás}}]{Gutschow10}%
  \BibitemOpen
  \bibfield  {author} {\bibinfo {author} {\bibfnamefont {J.}~\bibnamefont
  {Gütschow}}, \bibinfo {author} {\bibfnamefont {S.}~\bibnamefont {Uphoff}},
  \bibinfo {author} {\bibfnamefont {R.~F.}\ \bibnamefont {Werner}}, \ and\
  \bibinfo {author} {\bibfnamefont {Z.}~\bibnamefont {Zimborás}},\ }\href
  {\doibase 10.1063/1.3278513} {\bibfield  {journal} {\bibinfo  {journal}
  {Journal of Mathematical Physics}\ }\textbf {\bibinfo {volume} {51}},\
  \bibinfo {pages} {015203} (\bibinfo {year} {2010})},\ \bibinfo {note} {arXiv:
  0906.3195}\BibitemShut {NoStop}%
\bibitem [{\citenamefont {Kitaev}(2003)}]{Kitaev03}%
  \BibitemOpen
  \bibfield  {author} {\bibinfo {author} {\bibfnamefont {A.~Y.}\ \bibnamefont
  {Kitaev}},\ }\href@noop {} {\bibfield  {journal} {\bibinfo  {journal} {Annals
  of Physics}\ }\textbf {\bibinfo {volume} {303}},\ \bibinfo {pages} {2}
  (\bibinfo {year} {2003})}\BibitemShut {NoStop}%
\bibitem [{\citenamefont {Pais}(1966)}]{PaisRMP}%
  \BibitemOpen
  \bibfield  {author} {\bibinfo {author} {\bibfnamefont {A.}~\bibnamefont
  {Pais}},\ }\href {\doibase 10.1103/RevModPhys.38.215} {\bibfield  {journal}
  {\bibinfo  {journal} {Rev. Mod. Phys.}\ }\textbf {\bibinfo {volume} {38}},\
  \bibinfo {pages} {215} (\bibinfo {year} {1966})}\BibitemShut {NoStop}%
\end{thebibliography}%
\end{document}